\documentclass[twocolumn, notitlepage, superscriptaddress, aps]{revtex4-1}
\usepackage{xcolor}
\usepackage{amsmath}
\usepackage{fancyhdr}
\fancypagestyle{plain}{
\fancyfoot[C]{\thepage}}
\usepackage{amsthm}
\usepackage{graphicx}
\usepackage{float}
\usepackage{hyperref}
\usepackage{ulem}
\usepackage{upgreek}
\usepackage{caption}
\usepackage[position=top]{subfig}
\usepackage{framed}
\graphicspath{{Figures/}}

\newcommand{\ket}[1]{\left| #1 \right\rangle}
\newcommand{\bra}[1]{\left\langle #1 \right|}

\newcommand{\supp}[1]{(See Supplementary Section #1)}
\newcommand{\dg}{^\dagger}
\newcommand{\omg}{\omega}
\renewcommand{\v}[1]{\ensuremath{\mathbf{#1}}}
\newcommand{\od}[1]{^{(#1)}}

\newcommand{\gate}{\text{Gate}}
\newcommand{\pair}{\ket{\text{pair}}}
\newcommand{\bpair}{\bra{\text{pair}}}
\newcommand{\coded}{\ket{\text{coded}}}
\renewcommand{\subsection}[1]{\noindent \textbf{#1}\newline}

\begin{document}
\title{Joint measurement of time-frequency entanglement via sum frequency generation}
\author{Han Liu}
\affiliation{The Edward S.~Rogers Department of Electrical and Computer Engineering, University of Toronto, 10 King's College Road, Toronto, Ontario M5S 3G4, Canada}
\author{Amr S.~Helmy}
\email{a.helmy@utoronto.ca}
\affiliation{The Edward S.~Rogers Department of Electrical and Computer Engineering, University of Toronto, 10 King's College Road, Toronto, Ontario M5S 3G4, Canada}

\begin{abstract}
We propose, analyze, and evaluate a technique for the joint measurement of time-frequency entanglement between two photons. In particular, we show that the frequency sum and time difference of two photons could be simultaneously measured through the sum-frequency generation process, without measuring the time or frequency of each individual photon. We demonstrate the usefulness of this technique by using it to design a time-frequency entanglement based continuous variable superdense coding and a quantum illumination protocol. Performance analysis of these two protocols suggests that the joint measurement of strong time-frequency entanglement of non-classical photon pairs can significantly enhance the performance of joint-measurement based quantum communication and metrology protocols.
\end{abstract}

\maketitle
\section{Introduction}
Time and frequency correlation have been formidable resources in a rich range of applications from metrology and spectroscopy to communication and security \cite{liu2019enhancing,yabushita2004spectroscopy,lavoie2009quantum}.
In particular, time-frequency entanglement (TFE), that there are substantial and simultaneous correlations between two photons in both time and frequency, enables a plethora of advantages beyond what is achievable by classical correlations in the domains of metrology \cite{pe2005temporal} and communication\cite{mower2013high} applications. These advantages are brought about due to the continuous variable nature and loss resilient property\cite{zhang2008distribution} of TFE. Recent years have witnessed rapid advances in the generation of TFE with attractive properties \cite{horn2012monolithic,abolghasem2009bandwidth,svozilik2011generation}.
Thus far, TFE has already been used for entanglement based protocols such as quantum key distribution\cite{khan2006experimental,mower2013high} and continuous variable Bell test \cite{tittel1998violation}. Such protocols only feature separate measurements of the time or frequency degree of freedom of each individual photon. In contrast, many other entanglement based applications require a joint measurement of the two entangled photons, that is, measuring a joint variable of the entangled system without resolving the property of each individual photon. One such example is continuous variable superdense coding\cite{braunstein2000dense}, in which the sum of the in-phase amplitude \(x_\text{1},x_\text{2}\) and the difference of the out-of-phase amplitude \(p_\text{1}-p_\text{2}\) of two beams are simultaneous measured. In principle, such joint measurement based protocols could also be implemented with TFE photon pairs, since the time and frequency operator of a single photon obey the same commutation relation as \(x\) and \(p\)\cite{shalm2013three}.

Despite the progress on the front of TFE, to the best of our knowledge, TFE within non-classical photon pairs has not been used in quantum information and sensing applications that require joint measurement, where the sum of frequencies \(\omg_\text{1}+\omg_\text{2}\) and difference of times \(t_\text{1}-t_\text{2}\) of two photons are measured simultaneously. The frequency sum (time difference) of two photons has to be measured without measuring the frequency (time) of each individual photon, to avoid altering the subsequent measurement of the time difference (frequency sum). If such a measurement could indeed be implemented in a practical fashion, it could have a significant impact on many entanglement based applications that would have their performance depend on such joint measurements. Those include quantum teleportation\cite{furusawa1998unconditional}, quantum superdense coding\cite{bennett1992communication}, and quantum metrology \cite{lloyd2008enhanced}\cite{zhuang2017entanglement}.

Compared to the joint measurement, the generation of TFE is much more developed. To date, the most widely adopted approach to generating TFE photon pairs is continuous-wave spontaneous parametric down-conversion (SPDC). The TFE of SPDC photon pairs is closely related to the properties of SPDC sources. In particular, it has been shown that the temporal correlation and frequency anti-correlation of TFE photon pairs could be tailored with great flexibility with different designs of the waveguide structure of the photon pair source\cite{abolghasem2009bandwidth}. Given the close connection between the SPDC process and TFE, a natural line of inquiry would be to utilize the time-reversal of the SPDC process, namely sum frequency generation (SFG), to help obtain an effective route for a TFE joint-measurement based protocol.
In this paper, we show through theoretical analysis that the SFG process could be used as the joint measurement of TFE. In particular, we show that SFG could be used for the joint measurement of superdense coding and quantum illumination. These two examples show the potential of SFG as a measurement technique for practical quantum communication and sensing protocols.
\section{Results}
\subsection{Sum Frequency Generation of Photon Pairs}
The SFG process in a \(\chi\od{2}\) nonlinear medium could be modeled as the following evolution operator\cite{o2009time}:
\begin{gather}
V = I+\epsilon(\int d\omg_\text{p}d\omg_\text{s}d\omg_\text{i} f_0(\omg_\text{p},\omg_\text{s},\omg_\text{i})\nonumber\\
a_\text{p}\dg(\omg_\text{p})a_\text{s}(\omg_\text{s})a_\text{i}(\omg_\text{i})\delta(\omg_\text{p}-\omg_\text{s}-\omg_\text{i}) -H.C.)\label{H1}
\end{gather}
where photons in the \textit{signal} mode \(a_\text{s}(\omg_\text{s})\) and \textit{idler} mode \(a_\text{i}(\omg_\text{i})\) are annihilated to generate photons in the \textit{pump} mode \(a_\text{p}(\omg_\text{p})\)  and \(\epsilon\) characterizes the interaction strength.
The time reversed process of SFG, the SPDC process, can also be described by the same evolution operator \(V\). Note that the function \(f_0(\omg_\text{p},\omg_\text{s},\omg_\text{i})\delta(\omg_\text{p}-\omg_\text{s}-\omg_\text{i})\) is the joint spectral amplitude of SPDC photon pairs if the pump mode is occupied by strong coherent light at frequency \(\omg_\text{p}\). Given the fact that the SPDC process can create TFE photon pairs, it is natural to ponder whether the SFG process can be used to resolve the TFE between two photons. To investigate this while balancing the rigorousness and complexity of the analysis, we assume that the factor
\begin{gather}
f_0(\omg_\text{p},\omg_\text{s},\omg_\text{i}) = f(\frac{\omg_\text{s}-\omg_\text{i}}{\sqrt{2}})\label{eq2}
\end{gather}
is independent of the pump frequency \(\omg_\text{p}=\omg_\text{s}+\omg_\text{i}\). This indicates that the SPDC process is assumed to be broadband phase matched for any pump frequency.
Using this form of the SFG evolution operator \(V\) and work in the Heisenberg picture, the spectral density operator \(a\dg_\text{p}(\omg)a_\text{p}(\omg)\) of the pump light at the SFG output could be expressed as \supp{1.3}:
\begin{gather}
V\dg a\dg_\text{p}(\omg_\text{p})a_\text{p}(\omg_\text{p}) V
=\epsilon^2 B\dg B +O(a_\text{p}(\omg_\text{p}),a\dg_\text{p}(\omg_\text{p}))\label{HEI}\\
B = \iint d\omg_\text{s}d\omg_\text{i} f((\omg_\text{s}-\omg_\text{i})/\sqrt{2})\delta(\omg_\text{s}+\omg_\text{i}-\omg_\text{p}) a_\text{s}(\omg_\text{s})a_\text{i}(\omg_\text{i})\label{DEFB}
\end{gather}
where \(O(a_\text{p}(\omg_\text{p}),a\dg_\text{p}(\omg_\text{p}))\) is a sum of normal ordered operators that are at least linear in \(a_\text{p}(\omg_\text{p})\) or \(a_\text{p}\dg(\omg_\text{p})\). This term could be neglected due to the absence of pump photons at the input of the SFG process.
It could be further shown that in the limit of infinite SPDC photon bandwidth (\(f((\omg_\text{s}-\omg_\text{i})/\sqrt{2})=1\)):
\begin{gather}
P(\omg_\text{p},0) = \frac{1}{2\uppi}  B\dg B \label{prop}
\end{gather}
where \(P(\omg_\text{p},0)\) is the probability density operator for the input signal-idler photon pair of the SFG process to have their frequency sum \(\omg_\text{s}+\omg_\text{i}=\omg_\text{p}\) and zero time difference \(t_\text{s}-t_\text{i}=0\), simultaneously (see the Methods section). The general probability density operator \(P(\omg_\text{p},t)\) for the frequency sum \(\omg_\text{s}+\omg_\text{i}=\omg_\text{p}\) and time difference \(t_\text{s}-t_\text{i}=t\) could be constructed from \(P(\omg_\text{p},0)\) by temporally displacing either the signal or the idler photon. Therefore \eqref{HEI} and \eqref{prop} show that frequency resolved detection (i.e. using a single photon spectrometer, hence carrying out a classical measurement) of a pump photon generated in the SFG process reveals the simultaneous time and frequency correlation (hence TFE) \(\omg_\text{p}=\omg_\text{s}+\omg_\text{i},t_\text{s}-t_\text{i}=0\) of the input signal-idler photon pair, in the limit of infinite SPDC photon bandwidth. Intuitively, such a joint measurement of TFE could be understood as a quantum interference effect: only photon pairs that have frequency sum \(\omg_\text{p}=\omg_\text{s}+\omg_\text{i}\) can possibly generate a pump photon at frequency \(\omg_\text{p}\) due to the energy conservation constraint. Meanwhile in the time domain, a non-zero time difference between the signal and idler photon will induce different phase shift for different frequency components of the photon pair state, and the corresponding probability amplitude of SFG will interfere destructively, leading to a decreased probability of generating a pump photon.\\

\subsection{Time Frequency Entanglement Based Continuous Variable Superdense Coding}
Having shown that the SFG process could be utilized for the joint measurement of TFE, the question remains now is that whether it could benefit any quantum communication and sensing application that needs joint measurement. An important example is continuous variable superdense coding\cite{braunstein2000dense}, which utilizes entanglement between two particles to surpass the classical limit of channel information capacity. The previous proposal \cite{braunstein2000dense} and implementation\cite{li2002quantum} of continuous variable superdense coding are based on the joint-measurement of quadrature-phase entanglement, compared to which the joint measurement of TFE is more difficult to implement. However, the utilization of TFE can also provide additional advantages compared to the quadrature-phase entanglement. First, unlike the quadrature-phase entanglement, the strength of TFE (the Schmidt number of the photon pair) is not limited by the source power, which may translate to a larger information capacity enhancement compared to the two-fold enhancement achievable by quadrature-phase entanglement\cite{braunstein2000dense}. Second, the TFE has been demonstrated to be resilient to channel losses\cite{zhang2008distribution}, which is favourable for practical long haul superdense coding applications.

In what follows, we propose a proof-of-principle protocol of TFE based continuous variable superdense coding (TFE SDC). In particular, we will show that one can encode (decode) an arbitrarily large amount of information into (out of) both the time and frequency degree of freedom of the signal photon (that is entangled with the idler photon), simultaneously.  The basic steps of the TFE SDC protocol are as follows. First Alice generates an entangled photon pair state \(\pair\) as the entanglement source:
\begin{gather}
\pair = \int_{-\infty}^{+\infty}d\omg_\text{s}d\omg_\text{i} \phi_0^*(\omg_\text{s},\omg_\text{i})a_\text{s}\dg(\omg_\text{s})a_\text{i}\dg(\omg_\text{i})\ket{0}\\
\phi_0(\omg_\text{s},\omg_\text{i}) = h(\frac{\omg_\text{s}+\omg_\text{i}}{\sqrt{2}})f(\frac{\omg_\text{s}-\omg_\text{i}}{\sqrt{2}})\label{JSA}
\end{gather}
Such a photon pair could be generated by pumping the \(\chi\od{2}\) medium with a strong coherent beam of light in pump mode mode (\(a_\text{p}(\omg_\text{p})\)) that has (square normalized) complex spectral amplitude \(\frac{1}{\sqrt[4]{2}}h(\frac{\omg_\text{p}}{\sqrt{2}})\). To simplify the calculation, we assume the factor \(h(\omg)\) and \(f((\omg_\text{s}-\omg_\text{i})/\sqrt{2})\) to be Gaussian:
\begin{gather}
\frac{1}{\sqrt[4]{2}}h(\frac{\omg}{\sqrt{2}}) =  \sqrt{\frac{1}{\sqrt{2\uppi}\sigma_+}\exp(-\frac{(\omg-\omg_0)^2}{2\sigma_+^2})}\\
f(\omg) = \sqrt{\frac{1}{\sqrt{2\uppi}\sigma_-}\exp(-\frac{\omg^2}{2\sigma_-^2})}
\end{gather}
where the SPDC pump bandwidth and SPDC photon bandwidth are proportional to \(\sigma_+\) and \(\sigma_-\), respectively and \(\omg_0\) is the center frequency of the SPDC pump light. Alice then stores the idler photon locally and sends the signal photon to Bob. Bob will encode information by shifting both the frequency and time of the signal photon by \(\Updelta\upomega\) and \(\Updelta\text{t}\), which could be done with nonlinear frequency conversion\cite{kobayashi2016frequency} and a tunable delay line. The encoded signal photon is sent back to Alice. Then the information coded photon pair state for Alice to measure is given by:
\begin{gather}
\coded = \int d\omg_\text{s}d\omg_\text{i} \phi_\text{coded}^*(\omg_\text{s},\omg_\text{i})a_\text{s}\dg(\omg_\text{s})a_\text{i}\dg(\omg_\text{i})\ket{0}\\
\phi_\text{coded}= \phi_0(\omg_\text{s}-\Updelta\upomega,\omg_\text{i})\exp(\text{i}\omg_\text{s}\Updelta\text{t})
\end{gather}
\newcommand{\SFG}{\ket{\text{SFG}}}
\newcommand{\bSFG}{\bra{\text{SFG}}}
Alice will perform SFG with the encoded photon pair to obtain the final state \(\SFG\):

\begin{gather}
\SFG=V\coded
\end{gather}
The generated pump photon in state \(\SFG\) is sent to a single photon spectrometer. The frequency spectrum \(S(\omg)\) of the generated pump photon is given by the expectation value of the spectral density operator \(a_\text{p}\dg(\omg_\text{p}) a_\text{p}(\omg_\text{p})\)\supp{2}:
\begin{gather}
S(\omg_\text{p}) = \bSFG a_\text{p}\dg(\omg_\text{p})a_\text{p}(\omg_\text{p}) \SFG\\
=\frac{\epsilon ^2 \exp \left(\frac{1}{8} \left(-4 \Updelta\text{t}^2 \sigma _-^2-\frac{\Updelta\upomega ^2}{\sigma _-^2}-\frac{4 (\Updelta\upomega +\omg_0-\omega_\text{p} )^2}{\sigma_+^2}\right)\right)}{2 \sqrt{\uppi } \sigma _+}
\end{gather}
The total probability of generating a pump photon is given by the integral of \(S(\omg_\text{p})\):
\begin{gather}
N_\text{SFG} = \int d\omg_\text{p}S(\omg_\text{p}) =\frac{\epsilon^2}{\sqrt{2}}\exp(-\frac{\Updelta \upomega^2}{8\sigma_-^2}-\frac{\sigma_-^2\Updelta\text{t}^2}{2})\label{SD1}
\end{gather}
The mean frequency and frequency variance of the generated pump photon are given by:
\begin{gather}
\bar{\omg}_\text{SFG} = \frac{1}{N_\text{SFG}}\int d\omg_\text{p} \omg_\text{p} S(\omg_\text{p}) =\omg_0+\Updelta\upomega\label{SD15}\\
\text{var}\{\omg_\text{SFG}\}= \frac{1}{N_\text{SFG}}\int d\omg_\text{p} \omg_\text{p}^2 S(\omg_\text{p})-\bar{\omg}_\text{SFG}^2 =\sigma_+^2\label{SD2}
\end{gather}
\begin{figure}
\includegraphics[width=\columnwidth]{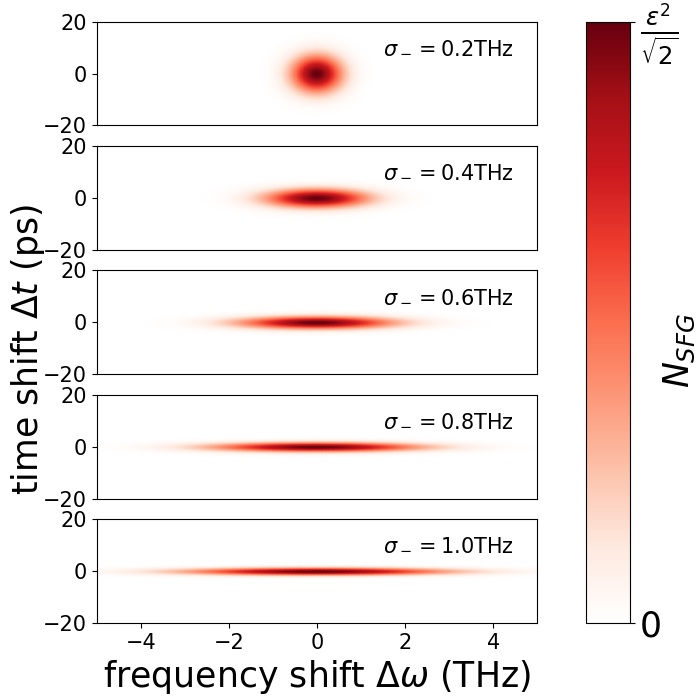}
\caption{\textbf{the probability of generating a pump photon} as a function of the frequency shift \(\Updelta\upomega\) and time shift \(\Updelta\text{t}\) for different SPDC photon bandwidth \(\sigma_-\) from \(0.2\)THz to \(1.0\)THz.\label{NSFG}}
\end{figure}
As could be seen in \eqref{SD1} and Fig. \ref{NSFG}, the probability \(N_\text{SFG}\) of generating a pump photon decreases rapidly as the signal photon time shift \(\Updelta\text{t}\) exceeds the inverse SPDC photon bandwidth \(1/\sigma_-\).  In contrast, the frequency shift \(\Updelta\upomega\) does not affect much \(N_\text{SFG}\) as long as \(\Updelta\upomega\ll\sigma_-\). The equation \eqref{SD15} and \eqref{SD2} show that the center frequency and bandwidth of the generated pump photon is identical to that of the SPDC pump light, aside from the frequency shift \(\Updelta\upomega\) that Bob encode into the photon pair.\\

\begin{figure}[h]
\centering
\includegraphics[width=\columnwidth]{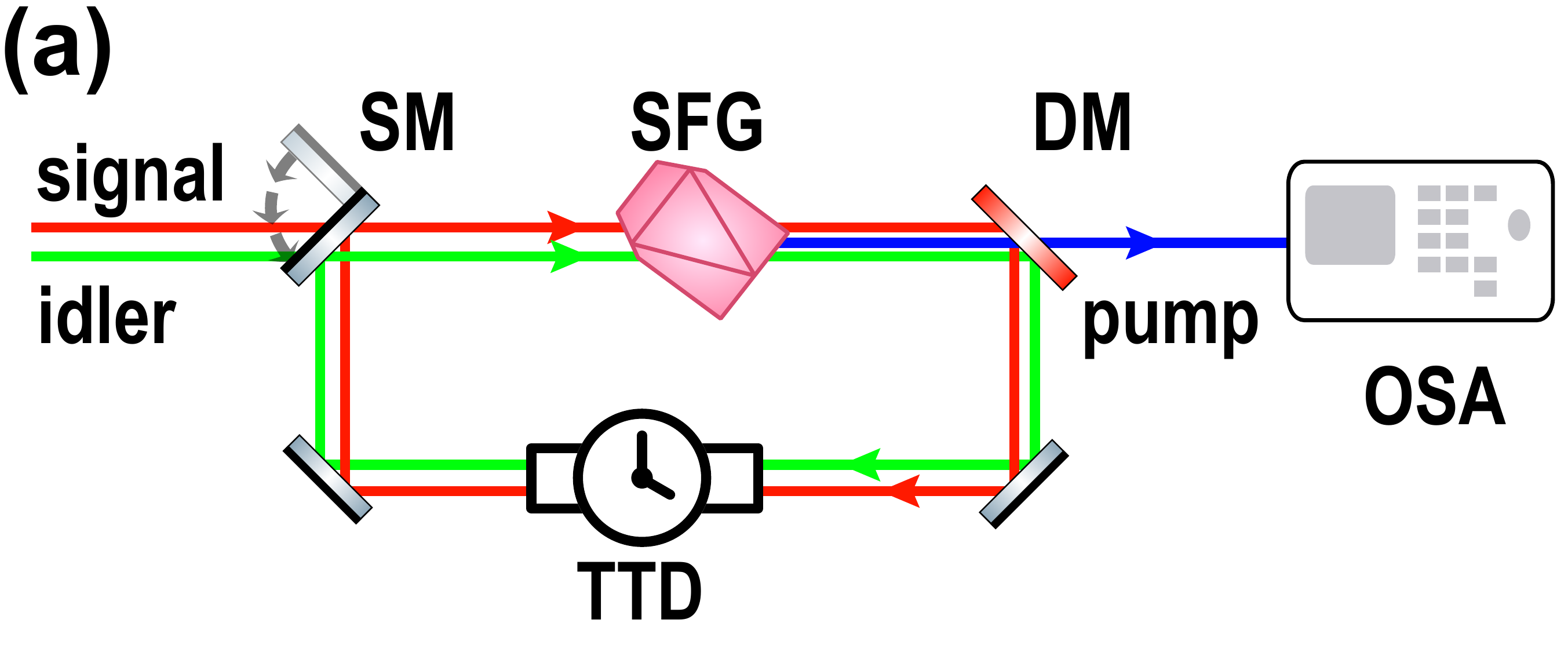}\\
\vspace{0.2cm}
\includegraphics[width=\columnwidth]{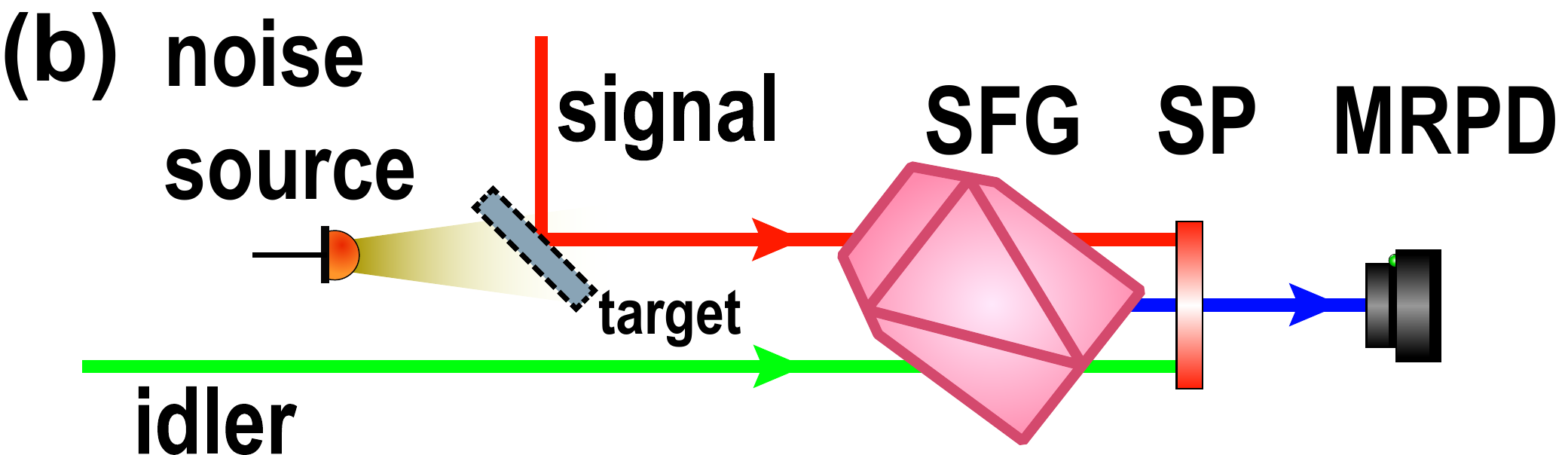}
\caption{\textbf{Experimental setup for the TFE SDC and TFE QI protocol.} (a): the setup of the SFG feedback loop for the TFE SDC protocol. SFG: a \(\chi\od{2}\) medium where the SFG process takes place; DM: a dichroic mirror to separate the generated pump photon from the photon pair; TTD: tunable time delay \(\Updelta\text{t}_\text{extra}\); OSA: single-photon optical spectral analyzer. SM: switch mirror. After the signal and idler pass by the switch mirror SM, SM will flip and form a ring cavity. (b): the setup of the TFE QI protocol. Target: the target object to be detected, modeled as a unballanced beamspliter with reflection \(\eta\) (\(\eta=0\) when the object is absent), SP: short pass filter, MRPD: mode resolved single photon detector that detects pump photon generated in mode \(A_0\).}\label{setup}
\end{figure}

Based on \eqref{SD1},\eqref{SD15} and \eqref{SD2}, the joint measurement scheme of the TFE SDC protocol could be designed as follows. After receiving the signal photon from Bob, Alice first apply an additional time shift \(\Updelta\text{t}_\text{extra}\) to the signal photon and then let the encoded photon pair go through the SFG process. A pump photon will be generated through SFG with non-negligible probability only if the total time shift \(\Updelta\text{t}+\Updelta\text{t}_\text{extra}\) is close to zero (\(\le1/\sigma_-\)). The encoded photon pair will remain unchanged after SFG if no pump photon is generated. In such cases, the encoded photon pairs could be reused and go through the SFG process repeatedly until a pump photon is finally generated (see Fig. \ref{setup} for the schematic of the experimental setup). Over this SFG feekback loop, the extra time delay \(\Updelta\text{t}_\text{extra}\) is swept continuously and repeatedly. When the feedback loop terminates (a pump photon is generated) with the extra time delay set to \(\Updelta\text{t}_\text{extra}\), the posterior probability distribution of the time shift \(\Updelta\text{t}\) is centered around \(-\Updelta\text{t}_\text{extra}\) with variance \({var}\{\Updelta\text{t}\}=1/\sigma_-^2\). The frequency shift \(\Updelta\upomega\) is obtained by measuring the frequency of the generated pump photon, with variance \({var}\{\Updelta\upomega\}=\sigma_+^2\). The measurement of \(\Updelta\text{t}\) and \(\Updelta\upomega\) could be arbitrarily precise simultaneously:
\begin{gather}
{var}\{\Updelta\text{t}\}{var}\{\Updelta\upomega\}=\sigma_+^2/\sigma_-^2\ll1
\end{gather}
which implies that arbitrarily large amount of information could be coded in \(\Updelta\text{t}\) and \(\Updelta\upomega\) simultaneously. Such information capacity cannot be achieved without using entanglement. To see this, consider a classical coding protocol where the time shift \(\Updelta\upomega\) and frequency shift \(\Updelta\text{t}\) information is coded on a single photon. Then the readout of \(\Updelta\text{t}\) and \(\Updelta\upomega\) information can only be done through measuring the time and frequency of the encoded photon. But the time and frequency of a single photon cannot be simultaneously measured with arbitrarily high accuracy due to the uncertainty principle \cite{qi2006single}.\\

As could be seen in \eqref{SD1} and \eqref{SD2}, the performance of the TFE SDC protocol depends on the \(\chi\od{2}\) medium that is used for the SPDC photon pair generation and the SFG measurement: the SPDC photon bandwidth \(\sigma_-\) dictates the maximal frequency shift \(\Updelta\upomega\) that can be encoded such that \(N_\text{SFG}\) is constantly \(\epsilon^2/\sqrt{2}\) as well as the readout variance of the time shift \(\Updelta\text{t}\). The nonlinear conversion efficiency \(\epsilon^2\) determines the number of the SFG feedback loops that are needed for the joint measurement, hence the speed of the communication. This outcome demonstrates how TFE could be used for superdense coding, where it offers advantages over existing superdense proposals and demonstrations in that it pivots in its performance on the TFE of the photon pairs, which is not limited by the power of the source, and is resilient to losses.\\

\subsection{Time Frequency Entanglement Based Quantum Illumination}
While advantages offered by TFE to superdense coding can clearly benefit channel capacity in communication systems, they can also benefit a number of sensing applications. In this work we aim to explore the possibility of utilizing TFE to enhance the target detection sensitivity in a noisy and lossy environment, i.e. the quantum illumination protocol. The quantum illumination  protocol is closely related to the superdense coding protocol in that they are both based on the same fashion of quantum entanglement enhancement\cite{lloyd2008enhanced}. The basic steps of the general quantum illumination protocol are as follows (see Fig. 1): The signal photon of the entangled photon pair is sent to probe the target while the idler photon is stored locally. The signal photon is reflected on the target object and collected with total transmission \(\eta\) (\(\eta=0\) if the target is absent). Regardless of the presence or absence of the target object, a constant level of environmental noise light is also always collected into the detection system. The joint measurement of the collected (signal or noise) photon and the idler photon is performed to predict the presence and absence of the target object. For TFE based quantum illumination protocol, the joint measurement consists of the sum frequency generation process of the idler photon and the collected photon, and the detection of the generated pump photon at the sum frequency \(\omg_\text{p} = \omg_\text{s}+\omg_\text{i}\). The entanglement enhancement provided by TFE results from the fact that only the signal photon that is strongly correlated in both time and frequency (hence TFE, which is not possible for uncorrelated photon pairs) with the idler photon can generate a pump photon at frequency \(\omg_\text{p}=\omg_\text{s}+\omg_\text{i}\). Therefore the detection of the generated pump photon is resilient to the background noise photons that are not correlated with the idler photon. Note that this protocol is entanglement based and is different from the target detection protocol using classical time and frequency correlation\cite{zhang2020multidimensional}. In what follows, we analyze the TFE QI protocol using a new mathematical formalism instead of that used for the superdense coding protocol. The formalism utilized here is chosen as it better highlights the connection between TFE and SFG (an analysis of the TFE QI protocol that parallels the analysis of the TFE SDC protocol could be found in Supplementary Section 5).

We shall start directly from the general form of the evolution operator \(V\) \eqref{H1} without applying any approximation or assumption. In general, the \(\chi\od{2}\) evolution operator \(V\) could be expressed as a discrete sum through a `two-step Schmidt decomposition process'(see the Methods section):
\begin{gather}
V = I+\epsilon\sum\limits_m(\sqrt{\lambda_m\od{1}} A_m\dg B_m -H.C.)\label{DH1}\\
B_m = \sum\limits_n \sqrt{\lambda_{m,n}\od{2}}F_{m,n}G_{m,n}\label{DH2}
\end{gather}
The equation \eqref{DH1} is obtained through the Schmidt decomposition between the pump and the ``signal-idler'' joint system with the singular values given by \(\{\lambda_m\od{1}\}\). The equation \eqref{DH2} is obtained through the Schmidt decomposition of each ``signal-idler'' joint system with the singular values given by \(\{\lambda_{m,n}\od{2}\}\). The operators \(\{A_m\}\) and \(\{F_{m,n}\}\),\(\{G_{m,n}\}\) (with fixed \(m\) and different \(n\)) form complete orthogonal sets of annihilation operators for the pump, signal and idler mode, respectively. The definitions of the mode operators could be found in the Methods section. The photon pair source of the TFE QI protocol is chosen to be \(\pair = B_0\dg\ket{0}\), which could be approximated by SPDC twin beams generated by coherent pump light in the mode \(A_0\)(neglecting the vacuum term and multiple pair terms). Note that if condition \eqref{eq2} is satisfied, then the mode \(A_0\) can be specified arbitrarily and \(\lambda_m\od{1}=1/\sqrt{2}\) \supp{3}. The noise and loss of the signal photon in the target detection channel is modeled as mixing with the background noise mode on a virtual beam-splitter with transmission \(\eta\) for the signal photons. The evolution operator \(U_\text{loss}\) of the beam-splitter can be expressed as:
\begin{gather}
U_\text{loss} = \prod\limits_n\exp\{\text{i}\arccos(\eta)(F_{0,n}\dg F\od{\text{b}}_{0,n}+H.C.)\}
\end{gather}
where \(F\od{\text{b}}_{0,n}\) is the discrete mode operator for the noise photon that has the same spectral amplitude as the signal mode \(F_{0,n}\). Equivalence between \(U_\text{loss}\) and the usual beam-splitter transform is shown in Supplementary Section 4.1. To avoid technical complexities, we assume that the background noise mode is occupied by a noise state \(\rho_\text{b}\) that satisfies the following conditions:
\begin{gather}
\text{tr}\{F_{0,n'}^{(\text{b})\dagger}F_{0,n''}^{\text{(b)}}\rho_\text{b}\}=\delta_{n'n''}\mu_\text{b} \hspace{0.5cm}\text{tr}\{F_{0,n}\od{\text{b}}\rho_\text{b}\}=0
\end{gather}
where \(\mu_\text{b}\) is the average number of noise photons per mode. The above conditions mean that the noise photons are evenly distributed in every signal mode \(F_{0,n}\) with random phases and there is no coherence between each mode. It could be shown that such noise is broadband and continuous-wave white noise \supp{4.2}. The density operator \(\rho_\text{SFG}\) of the SFG output can be expressed as:
\begin{gather}
\rho_\text{SFG} = VU_\text{loss}\pair\bpair\otimes\rho_\text{b} U_\text{loss}\dg V\dg
\end{gather}
In the limit of perfect signal photon transmission (\(\eta=1\)), the SFG process can only generate pump photons in mode \(A_0\) \supp{4.3}. For this reason, in the following analysis only the photon detection event in mode \(A_0\) is taken into consideration. After some algebraic manipulation \supp{4.3}, it can be shown that the photon detection probability \(P_\text{d,QI}\) on mode \(A_0\) is given by:
\begin{gather}
P_\text{d,QI} = \text{tr}\{A_0\dg A_0 \rho_\text{SFG}\} = \epsilon^2\lambda_{0}\od{1}(\eta+\mu_\text{b}/\text{SN}) \label{PD}
\end{gather}
where \(\text{SN}=1/\sum_n(\lambda_{0,n}\od{2})^2\) is the Schmidt number of the SPDC photon pair. The Schmidt number being larger than unity is a indication of TFE. It can be shown that the conversion efficiency \(\lambda_0\od{1}\epsilon^2\) is also the SPDC conversion efficiency \supp{4.5}. For comparison, we shall also consider a classical target detection (CI) protocol where a probe photon in an arbitrary temporal-spectral mode \(F\) is sent to probe the target\cite{lloyd2008enhanced}. It is easy to see that the photon detection probability is \(P_\text{d,CI} = \eta+\mu_\text{b}\). Comparison between \(P_\text{d,CI}\) and \(P_\text{d,QI}\) shows that the TFE QI protocol is equivalent to a CI protocol with detection efficiency \(\epsilon^2\lambda_{0}\od{1}\) and noise photon per mode reduced to \(\mu_\text{b}/\text{SN}\).

As could be seen in \eqref{PD}, the performance of the TFE QI protocol is limited by the nonlinear efficiency \(\lambda_0\od{2}\epsilon^2\) and the Schmidt number SN of the entangled photon pair source. If SPDC twin beams are used as the TFE QI source, the Schmidt number SN could be approximated as the ratio of the SPDC photon bandwidth \(\sigma_-\) and the SPDC pump bandwidth \(\sigma_+\)\cite{bogdanov2007schmidt}. Therefore ideally the \(\chi\od{2}\) medium should have very large phase-matching bandwidth. For bulk \(\chi\od{2}\) crystal there is a trade-off between the SPDC photon bandwidth and the length of the crystal(hence the nonlinear conversion efficiency). Therefore it may not be optimal for the TFE QI protocol. Integrated semiconductor \(\chi\od{2}\)  waveguide \cite{horn2012monolithic} could be an ideal alternative because it offers high nonlinear conversion efficiency in a compact form factor (\(\lambda_0\od{1}\epsilon^2\simeq2.1\times10^{-8}\) for a 1mm long waveguide). Moreover, semiconductor waveguide can provide very large SPDC photon bandwidth with specific structure designs \cite{abolghasem2009bandwidth}.\\

It is important to highlight that the noise reduction being directly proportional to the Schmidt number SN is very similar to that in the first quantum illumination protocol reported in \cite{lloyd2008enhanced}. Therefore the TFE QI protocol proposed here could be considered as an implementation of \cite{lloyd2008enhanced}. However, it could be shown that the TFE QI protocol has large performance enhancement over the coherent light/homodyne detection scheme under high noise condition \supp{6.2}. This extends the result in \cite{shapiro2009quantum} that shows that the first quantum illumination protocol \cite{lloyd2008enhanced} cannot outperform coherent detection in the low noise limit \(\mu_\text{b}\ll1\). In addition, as could be seen in \eqref{PD}, the entanglement enhancement of the TFE QI protocol can effectively reduce the environmental noise power down to zero in the limit of large entanglement \(\text{SN}\gg1\). Such a result does not contradict the previous finding\cite{tan2008quantum} that at most 6dB of performance advantage could be achieved by Gaussian state quantum illumination. This is because for TFE QI the photon-pair source is assumed to be non-Gaussian \(\pair=B_0\dg\ket{0}\). However, in practice, the SPDC twin beams are commonly used as an approximation of \(\pair\) by neglecting the vacuum term and the multiple pair terms. Then, it must be remembered that SPDC twin-beams are in Gaussian state and a TFE QI protocol with SPDC twin-beam source can achieve 6 dB enhancement of target detection performance at most. Lastly, the proposed TFE QI protocol is similar to the SFG quantum illumination protocol\cite{zhuang2017optimum} but with a simpler setup. However, the discussion of the TFE QI protocol here provides a different perspective of the performance advantage achievable by SFG detection from a TFE standpoint.  \\
\section{DISCUSSION}
In this paper, we propose a technique to jointly measure the time-frequency entanglement between two photons based on the sum frequency generation process. We also apply this technique to propose a time-frequency entanglement based continuous variable superdense coding protocol and a quantum illumination protocol and analyze their performances. In particular, a theoretical analysis of the quantum illumination protocol shows that the effect of background noise on target detection accuracy can be reduced to zero in the limit of infinite time-frequency entanglement between the signal and idler photon. The performance limiting factor of this joint measurement technique is that its efficiency is limited by the strength of \(\chi\od{2}\) nonlinearity. To overcome this limit for the superdense coding protocol, we propose a feedback loop based setup to effectively enhance the interaction of the signal and idler photon inside the nonlinear medium. Other approaches to improving the efficiency may include resonance enhanced sum frequency generation of entangled photon pairs\cite{sensarn2009resonant} and enhancing the effective nonlinearity with high confinement waveguide design\cite{horn2012monolithic}.
\section{METHODS}
\newcommand{\ws}{{\omg_\text{s}}}
\newcommand{\wi}{{\omg_\text{i}}}
\subsection{The probability density operator \(P(\omg,t)\) for time difference and frequency sum of two photons}
Define first the frequency sum projection operator \(P_{\delta\omg}(\omg)\) that selects two photon states with the frequency sum of the signal and idler photon satisfying \(|\omg_\text{s}+\omg_\text{i}-\omg|\le\delta\omg/2\):
\begin{gather}
P_{\delta\omg}(\omg)\nonumber\\
= \iint d\ws d\wi a_\text{s}\dg(\ws)a_\text{i}\dg(\wi) a_\text{s}(\ws)a_\text{i}(\wi)\gate(\frac{\omg-\ws-\wi}{\delta\omg})
\end{gather}
where \(\gate(x)=1\) for \(|x|\le1/2\) and \(\gate(x)=0\) otherwise. The time difference projection operator that selects two photon states with the time difference of the signal and idler photon satisfying \(|t_\text{s}+t_\text{i}-t|\le\delta t/2\) can be similarly defined as:
\begin{gather}
P_{\delta t}(t)\nonumber\\
= \iint dt_\text{s} dt_\text{i} \tilde{a}_\text{s}\dg(t_\text{s})\tilde{a}_\text{i}\dg(t_\text{i}) \tilde{a}_\text{s}(t_\text{s})\tilde{a}_\text{i}(t_\text{i})\gate(\frac{t_\text{s}-t_\text{i}-t}{\delta t})
\end{gather}
where the instantaneous annihilation operator \(\tilde{a}_x(t)\) is defined as:
\begin{gather}
\tilde{a}_x(t) = \frac{1}{\sqrt{2\pi}}\int d\omg \exp(-i\omg t) a_x(\omg) \hspace{0.5cm}(x=\text{s,i})
\end{gather}
The probability density operator \(P(\omg,t)\) for time difference and frequency sum of two photons can then be defined as the product of \(P_{\delta t}(t)\) and \(P_{\delta\omg}(\omg)\) in the limit of \(\delta\omg\to0,\delta t\to0\):
\begin{gather}
P(\omg,t) = \lim\limits_{\delta t\to0,\delta\omg\to0}\frac{1}{\delta\omg\delta t} P_{\delta\omg}(\omg) P_{\delta t}(t)
\end{gather}
It can be shown that \supp{1.2}
\begin{gather}
P(\omg,t) =\frac{1}{2\pi} B_\text{p}\dg B_\text{p}
\end{gather}
where
\begin{gather}
B_\text{p}=\iint d\ws d\wi \delta(\ws+\wi-\omg)\exp(i\wi t)a_\text{s}(\ws)a_\text{i}(\wi)
\end{gather}
As can be seen \(B_\text{p}\) equals \(B\) defined in \eqref{DEFB} in the limit of \(f((\omg_\text{s}-\omg_\text{i})/\sqrt{2})\to 1\) and \(t=0\).\\

\subsection{The two step Schmidt decomposition}
To simplify the analysis of the SFG process, the evolution operator could be discretized from the integral form \eqref{H1} to a discrete sum form via a `two-step Schmidt decomposition'. First, the function \(\delta(\omg_\text{p}-\omg_\text{s}-\omg_\text{i})f_0(\omg_\text{p},\omg_\text{s},\omg_\text{i})\) could be decomposed through the first step Schmidt decomposition:
\begin{gather}
\delta(\omg_\text{p}-\omg_\text{s}-\omg_\text{i})f_0(\omg_\text{p},\omg_\text{s},\omg_\text{i})\nonumber\\
=\sum\limits_m \sqrt{\lambda_m\od{1}} \psi_{\text{A},m}^*(\omg_\text{p}) \psi_{\text{B},m}(\omg_\text{s},\omg_\text{i})
\end{gather}
where \(\{\psi_{\text{A},m}(\omg_\text{p})\}\) is a complete set of orthonormal functions for the complex amplitude of the pump photons and \(\{\psi_{\text{B},m}(\omg_\text{s},\omg_\text{i})\}\) is a (not complete) set of  orthonormal functions for the joint spectral amplitude of the signal and idler photon pairs. Then the function \(\psi_{\text{B},m}(\omg_\text{s},\omg_\text{i})\) could be further decomposed through the second step Schmidt decomposition:
\begin{gather}
\psi_{\text{B},m}(\omg_\text{s},\omg_\text{i})\nonumber\\
= \sum\limits_n \sqrt{\lambda_{m,n}\od{2}}\psi_{\text{F},m,n}(\omg_\text{s})\psi_{\text{G},m,n}(\omg_\text{i})
\end{gather}
where \(\{\psi_{\text{F},m,n}(\omg_\text{s})\}\)(for fixed \(m\)) and \(\{\psi_{\text{G},m,n}(\omg_\text{i})\}\)(for fixed \(m\)) are two complete sets of orthonormal functions for the signal and idler photon complex spectral amplitude, respectively. Therefore the evolution operator \(V\) could be written in a discrete sum form:
\begin{gather}
V = I+\epsilon\sum\limits_m(\sqrt{\lambda_m\od{1}} A\dg_m B_m -H.C.)\\
B_m = \sum\limits_n \sqrt{\lambda_{m,n}\od{2}}F_{m,n}G_{m,n}\\
A_m = \int d\omg \psi_{\text{A},m}(\omg) a_\text{p}(\omg)\\
B_m = \int d\omg_\text{s} d\omg_\text{i} \psi_{\text{B},m}(\omg_\text{s},\omg_\text{i}) a_\text{s}(\omg_\text{s})a_\text{i}(\omg_\text{i})\\
F_{m,n} = \int d\omg \psi_{\text{F},m,n}(\omg) a_\text{s}(\omg)\\
G_{m,n} = \int d\omg \psi_{\text{G},m,n}(\omg) a_\text{i}(\omg)
\end{gather}
\section*{Data availability}
Data sharing not applicable to this article as no datasets were generated or analyzed during the current study.
\section*{COMPETING INTERESTS}
The authors declare no competing interests.
\section*{Author contribution}
L.H. carried out the derivation and calculation of the result with assistance of A.H. Both authors contributed to writing the paper.
\section*{Figure Legends}
\textbf{the probability of generating a pump photon as a function of the frequency shift \(\Updelta\upomega\) and time shift \(\Updelta\text{t}\)} for different SPDC photon bandwidth \(\sigma_-\) from \(0.2\)THz to \(1.0\)THz.\\

\textbf{Experimental setup for the TFE SDC and TFE QI protocol} (a): the setup of the SFG feedback loop for the TFE SDC protocol. SFG: a \(\chi\od{2}\) medium where the SFG process takes place; DM: a dichroic mirror to separate the generated pump photon from the photon pair; TTD: tunable time delay \(\Updelta\text{t}_\text{extra}\); OSA: single-photon optical spectral analyzer. SM: switch mirror. After the signal and idler pass by the switch mirror SM, SM will flip and form a ring cavity. (b): the setup of the TFE QI protocol. Target: the target object to be detected, modeled as a unballanced beamspliter with reflection \(\eta\) (\(\eta=0\) when the object is absent), SP: short pass filter, MRPD: mode resolved single photon detector that detects pump photon generated in mode \(A_0\).
\bibliographystyle{unsrt}

\end{document}